\documentclass[sigconf]{acmart}
\usepackage{graphicx}
\usepackage{subcaption}
\usepackage{balance}  
\usepackage{xcolor}
\usepackage{makecell}
\usepackage{hyperref}
\usepackage{adjustbox}
\usepackage{float}
\usepackage{circledsteps}

\AtBeginDocument{%
  \providecommand\BibTeX{{%
    \normalfont B\kern-0.5em{\scshape i\kern-0.25em b}\kern-0.8em\TeX}}}

\setcopyright{acmcopyright}
\copyrightyear{2018}
\acmYear{2018}
\acmDOI{10.1145/1122445.1122456}

\acmConference[Woodstock '18]{Woodstock '18: ACM Symposium on Neural
  Gaze Detection}{June 03--05, 2018}{Woodstock, NY}
\acmBooktitle{Woodstock '18: ACM Symposium on Neural Gaze Detection,
  June 03--05, 2018, Woodstock, NY}
\acmPrice{15.00}
\acmISBN{978-1-4503-XXXX-X/18/06}

\begin{document}

\title{ModelCI-e: Enabling Continual Learning in Deep Learning Serving Systems}


\author{Yizheng Huang}
\authornotemark[1]
\affiliation{\institution{Nanyang Technological University}}
\email{yizheng.huang@ntu.edu.sg}

\author{Huaizheng Zhang}
\authornote{Both authors contributed equally to this research.}
\affiliation{\institution{Nanyang Technological University}}
\email{huaizhen001@e.ntu.edu.sg}

\author{Yonggang Wen}
\affiliation{\institution{Nanyang Technological University}}
\email{ygwen@ntu.edu.sg}

\author{Peng Sun}
\affiliation{\institution{SenseTime Group Limited}}
\email{sunpeng1@sensetime.com}

\author{Nguyen Binh Duong TA}
\affiliation{\institution{Singapore Management University}}
\email{donta@smu.edu.sg}

\begin{abstract}

MLOps is about taking experimental ML models to production, i.e., serving the models to actual users. Unfortunately, existing ML serving systems do not adequately handle the dynamic environments in which online data diverges from offline training data, resulting in tedious model updating and deployment works. This paper implements a lightweight MLOps plugin, termed ModelCI-e (continuous integration and evolution), to address the issue. Specifically, it embraces continual learning (CL) and ML deployment techniques, providing end-to-end supports for model updating and validation without serving engine customization. ModelCI-e includes 1) a model factory that allows CL researchers to prototype and benchmark CL models with ease, 2) a CL backend to automate and orchestrate the model updating efficiently, and 3) a web interface for an ML team to manage CL service collaboratively. Our preliminary results demonstrate the usability of ModelCI-e, and indicate that eliminating the interference between model updating and inference workloads is crucial for higher system efficiency.

\end{abstract}




\maketitle

\section{Introduction}

\begin{table*}[ht]
\centering
\caption{Comparison of several continual learning (CL) framework.}
\label{tab:system_compare}
\begin{adjustbox}{max width=\textwidth}
\begin{tabular}{cccccccccc}
\hline
Project             & \makecell{Continuous \\ monitoring} & \makecell{Continuous \\ updating} & \makecell{Drift \\ detection} & \makecell{CL data \\ cache} & \makecell{Model history \\ management} & \makecell{CL algorithm \\ benchmark} & \makecell{Service \\ validations} &  \makecell{CL algorithm \\ zoo} & \makecell{Portability} \\ \hline
Continuum \cite{tian2018continuum}             &                       & \checkmark               &                 &               &                &                &                    &                & \checkmark     \\ \hline
AWS Sagemaker \cite{sagemaker}                & \checkmark            &                          & \checkmark      &               &                &                &                    &                &                \\ \hline
Avalanche \cite{lomonaco2021avalanche}                     &                       & \checkmark               &                 &               &                & \checkmark     &                    & \checkmark     &                \\ \hline
River \cite{2020river}                         &                       & \checkmark               & \checkmark      &               &                &                &                    &                &                \\ \hline
ModelCI-e                                   & \checkmark            & \checkmark               & \checkmark     & \checkmark     & \checkmark     & \checkmark     & \checkmark         & \checkmark     & \checkmark     \\ \hline
\end{tabular}
\end{adjustbox}
\end{table*}

Machine Learning (ML), especially Deep Learning (DL), has become the essential component of many applications, ranging from video analytics \cite{zhang2021serverless} to resume assessment \cite{luo2018resumenet}. Consequently, inference accounts for a large proportion of ML product costs \cite{infercost2, infercost3}. To reduce the inference costs, both industry and academia have invested a lot of efforts in developing high-performance DL serving systems (e.g., TensorFlow Serving \cite{olston2017tensorflow} and Clipper \cite{crankshaw2017clipper}). These systems streamline model deployment and provide optimizing mechanisms (e.g., batching) for higher efficiency, thus reducing inference costs.

We note that existing DL serving systems are not able to handle the dynamic production environments adequately. The main reason comes from the concept drift \cite{gama2014survey, klaise2020monitoring} issue - DL model quality is tied to offline training data and will be stale soon after deployed to serving systems. For instance, spam patterns keep changing to avoid detection by the DL-based anti-spam models. In another example, as fashion trends shift rapidly over time, a static model will result in an inappropriate product recommendation. If a serving system is not able to quickly adapt to the data shift, the quality of inference would degrade significantly. Therefore, engineers must track the system performance carefully and update the deployed DL models in time. This results in much tedious, manual work in the operations of DL serving systems.

Several attempts have been made to address the concept drift issue, as summarized in Table \ref{tab:system_compare}. These efforts generally consider the issue from an algorithm perspective; or the system perspective. The former primarily focuses on improving continual learning (CL), also as referred to lifelong learning \cite{chen2018lifelong} to maintain model accuracy, while the latter (e.g., AWS SageMaker \cite{sagemaker}) pays more attention to developing a serving platform equipped with tools such as performance monitors and drift detectors. However, 1) these solutions still require users to do cumbersome DL serving system customization works including development, validation, and deployment, 2) none of them can provide full support for delivering a high-quality CL service, which requires close collaboration between model designers and production (e.g., DevOps) engineers, etc., and 3) how to efficiently orchestrate both updating (training) and inference workload in an ML cluster has not been considered.


To fill the gaps, we propose a new system with the following design guidelines. First, the system should be easily integrated with existing serving systems. Second, the system should support the entire model updating workflow to foster team collaboration and automate most of the labor-intensive operations. Third, the system should be efficient when training and deploying CL in practice.

In this paper, we describe ModelCI-e (continuous integration \cite{zhang2020mlmodelci} and evolution), an automated and efficient plugin that can be easily integrated into existing model serving systems (e.g., TensorFlow Serving, Clipper). We also conduct some preliminary studies to illustrate the potential challenges (e.g., interference of training and inference jobs in a cluster) for the system as well as the future optimization directions. To the best of our knowledge, ModelCI-e is the first holistic plugin system that can benefit both DL system research and operations.

\section{System Design and Implementation}

\begin{figure}[ht]
  \centering
  \includegraphics[width=1.0\linewidth]{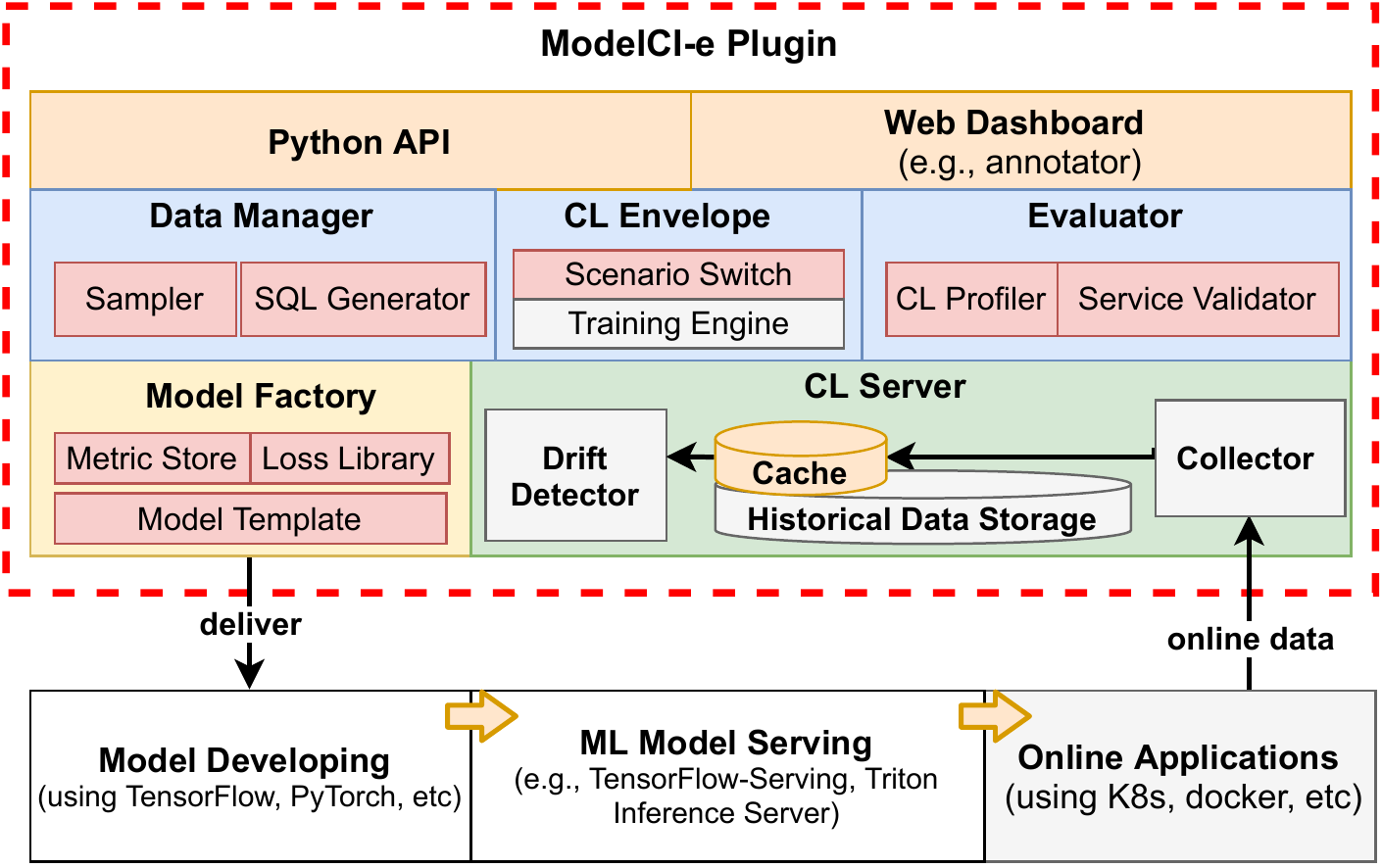}
  \caption{ModelCI-e architecture.}
  \label{fig:workflow}
\end{figure}

This section first summarizes the system workflow and then presents the functionalities and implementation details of the core components in ModelCI-e, as illustrated in Fig. \ref{fig:workflow}.

\subsection{Workflow}

The complete workflow of developing a CL-based ML service can be split into two phases, the offline preparation phase and the online orchestration phase. In the offline phase, researchers first leverage the built-in APIs of the model factory to customize a CL model with their favorite DL frameworks (e.g., PyTorch). The model is then tested and deployed to an existing serving engine like the Triton Inference Server \cite{triton} as a service by production engineers. Meanwhile, engineers can prepare a configuration file to integrate our system into the serving engine according to our provided template. This configuration file includes essential information to orchestrate the serving and the CL process. After these steps, a CL-based ML service is ready to start and process users' requests.

\begin{figure*}[ht]
  \centering
  \includegraphics[width=1.0\linewidth]{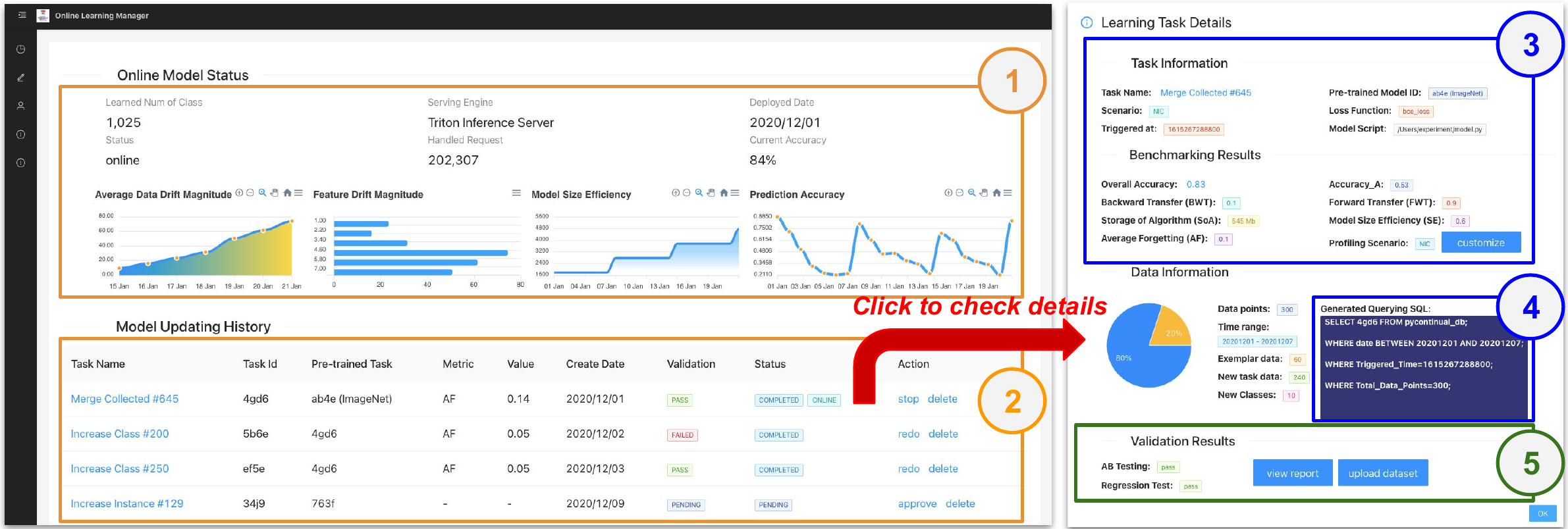}
  \caption{The dashboard of ModelCI-e. We define two kinds of users, including model researchers ((blue color) and production engineers (orange color for DevOps engineers and green color for quality assurance engineer), and put all of them into the same page.}
  \label{fig:panel}
\end{figure*}

In the online orchestration phase, the system follows the predefined rules in the configuration file to schedule model updating tasks while maintaining the quality of service (QoS). First, we launch a CL server in which 1) a drift detector is utilized to monitor model performance and decide when to trigger the model updating, and 2) a collector is used to aggregate requests and save them into a cache for the model updating purpose. Second, once a data drift exceeding the predefined threshold is detected, the system will invoke a suitable CL method as well as prepare the corresponding training samples to improve the model online. These operations can build a training job that will be dispatched to suitable workers. Third, once the updating is accomplished, our system will call a service validator that implements many off-the-shelf testing methods to verify the updated model. If the newly updated model passes all checks, the system will replace the existing models with it. The whole process needs no human involved, while the ML team can check and manage each step's output from the web interface.

\subsection{System Implementation}

\textbf{Model Factory} helps model researchers prototype and evaluate CL models quickly. It consists of three functions, model template, loss library, and metric store. (1) Model template is implemented atop PyTorch-lighting and contains high-level templates which make it easy to build popular CL applications (e.g., image and text recognition). (2) Loss library incorporates many popular CL loss functions like Elastic Weight Consolidation (EWC) \cite{kirkpatrick2017overcoming}, Synaptic Intelligence (SI) \cite{zenke2017continual}, etc. and provides Python APIs for researchers to upgrade their existing training loss functions to overcome the catastrophic forgetting \cite{mccloskey1989catastrophic} in CL. (3) Metric store helps users evaluate CL services from two perspectives, learning capability, and efficiency. The former includes metrics like BWT (backward transfer that measures the influences of a CL method on previous tasks) \cite{diaz2018don}, whereas the latter contains metrics such as computational efficiency, model size (MS) efficiency, etc.

\textbf{CL Server} is a lightweight Python service implemented with Flask and can be integrated with many serving engines (e.g., Tensorflow Serving). It implements 1) a data collector to collect and save online request data, which will be selectively utilized for human labeling and model upgrading, 2) an in-memory CL cache based on Redis to optimize the data sampling and loading efficiency, 3) a drift detector, including many detection methods (e.g., Kolmogorov-Smirnov \cite{dos2016fast} and EDDM \cite{baena2006early}), to calculates the data drift magnitude according to a predefined rule (e.g., calculation interval and accuracy requirement) in the configuration file and trigger the CL process, and 4) a hybrid orchestrator to automate the CL service update workflow for reducing manual work and to schedule jobs based on user-defined metrics (e.g., GPU utilization) for improving system efficiency.


\textbf{Data Manager} consists of three functions, which offer intuitive and convenient data management. Specifically, we provide 1) a data sampler for data rehearsal (a CL method), which samples training data from historical data and then mixes them with newly collected data for training, 2) a SQL query generator to automatically navigate utilized data samples so researchers can easily reproduce, debug and verify CL models, and 3) an annotator implemented using label-studio \cite{labelstudio} for data labeling.

\textbf{CL Envelope} written in Python, automates the model updating job in which an updating scenario (mode) will be selected, and a training engine (e.g., PyTorch, and TensorFlow) will run upon receiving the drift detector's start signal. First, for a CL task, selecting the appropriate training scenario is vital to success, so we implement a scenario switch to manage different scenarios including new class learning (NC), new instance learning (NI), new instance and class learning (NIC) \cite{lomonaco2017core50} and offline retraining. Users then can define rules to select a scenario to update their models. For instance, if the system detects that the dominant proportion of the newly collected data has many new classes, the training will switch to NC mode. Second, a training engine will aggregate the training-related components like the data loader and the CL model to implement an updating job with a unique job ID.

\textbf{Evaluator} has two functions, CL profiler, and service validator, aiming to provide detailed performance reports of CL services in the system. The profiler allows researchers and practitioners to benchmark their CL models while the service validator contains many validation methods (e.g., A/B test) to make sure that each updated model meets the user-defined standards before they go online. Both of them will save their results in a MongoDB database and display a report in our dashboard implemented with React.js.

\begin{figure}
  \centering
  \includegraphics[width=1.0\linewidth]{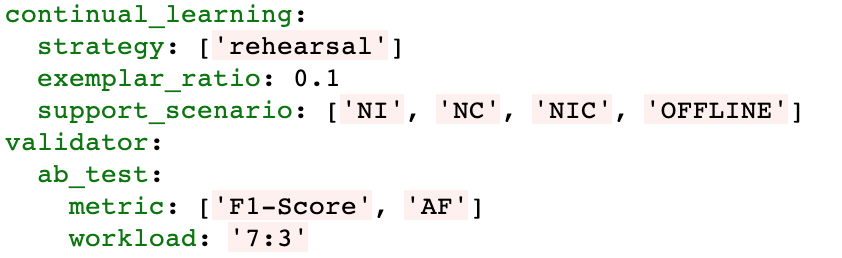}
  \caption{Part of the configuration file for developing a CL service.} 
  \label{fig:config}
\end{figure}

\section{Preliminary Results}



This section provides case studies that demonstrates ModelCI-e's usability and collaborations, presents the preliminary results about the system efficiency, and discusses the future research directions.

\textbf{Experiment Setup.} Unless otherwise noted, all experiments are conducted with the following settings. We employ widely deployed services like image recognition (IR) with ResNet50 and text classification (TC) with BERT, as evaluation workloads. For the former, we use the CORe50 \cite{lomonaco2017core50} dataset to simulate the continual learning (CL) process, whereas, for the latter, we use the IMDB Review dataset.  We deploy our system to an NVIDIA DGX station server with a 20-core E5-2698 v4 CPU and 4 Tesla V100 GPUs. 

\textbf{Evaluation 1: How does a team utilize ModelCI-e to enable efficient, continuous model updates in a typical ML serving system?} First, a model researchers utilize their domain knowledge and our model factory's APIs to set up a CL plugin for the IR service with a few lines of code. They also can evaluate the new CL-based IR service in our system. Next, the researchers will deliver the plugin to the production engineers, who will update the service configuration YAML file (as shown in Figure \ref{fig:config}) and integrate the plugin to the existing regular IR service. Finally, with a simple start command, the upgraded service will be validated and deployed online as shown in Figure \ref{fig:panel}. 

\begin{figure}
\centering     
\begin{subfigure}{0.235\textwidth}  
  \label{fig:image_data}
  \includegraphics[width=40mm]{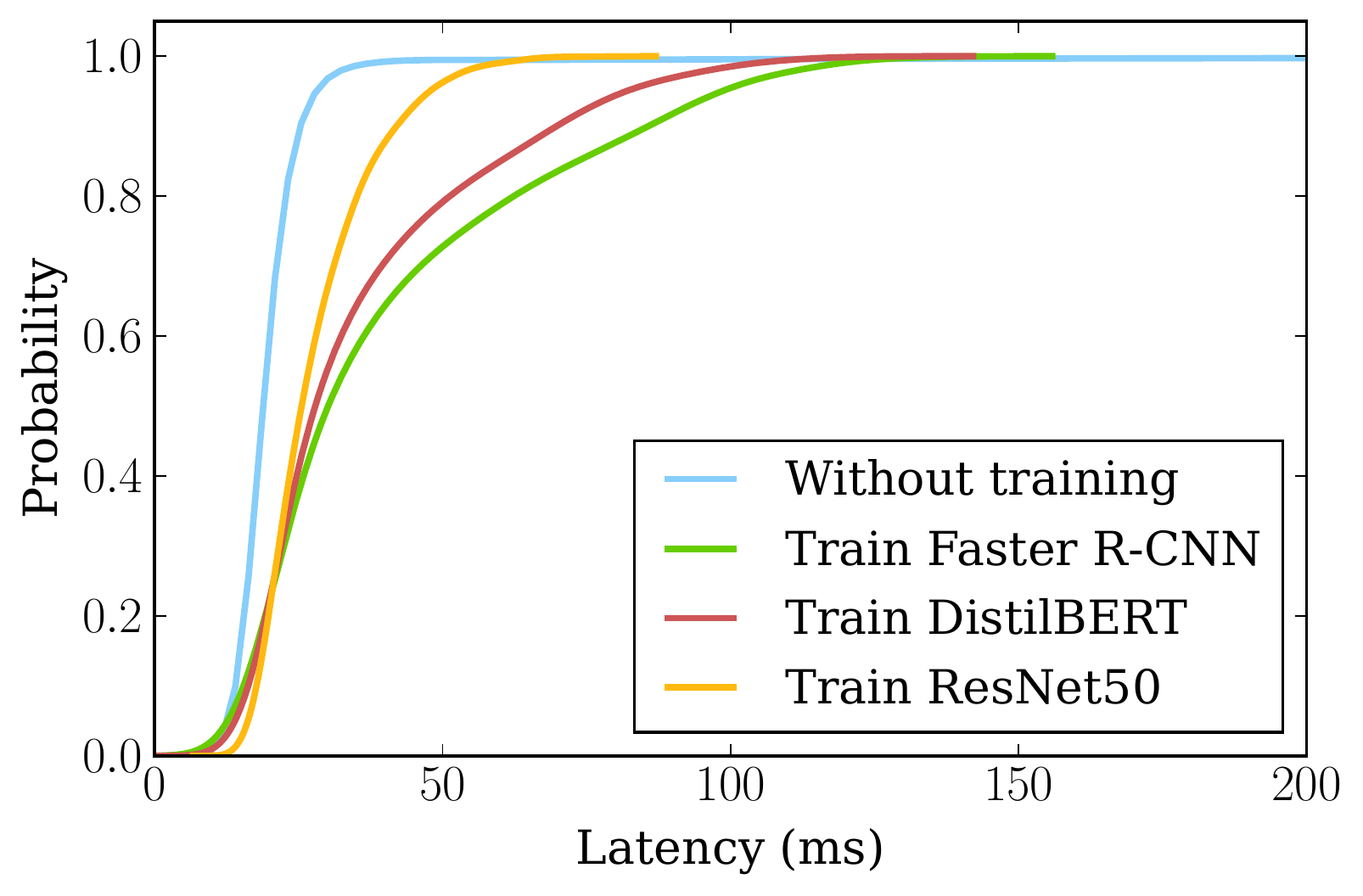} 
  \subcaption{Inference latency.}
\end{subfigure}
\begin{subfigure}{0.235\textwidth}  
    \label{fig:text_data}
    \includegraphics[width=40mm]{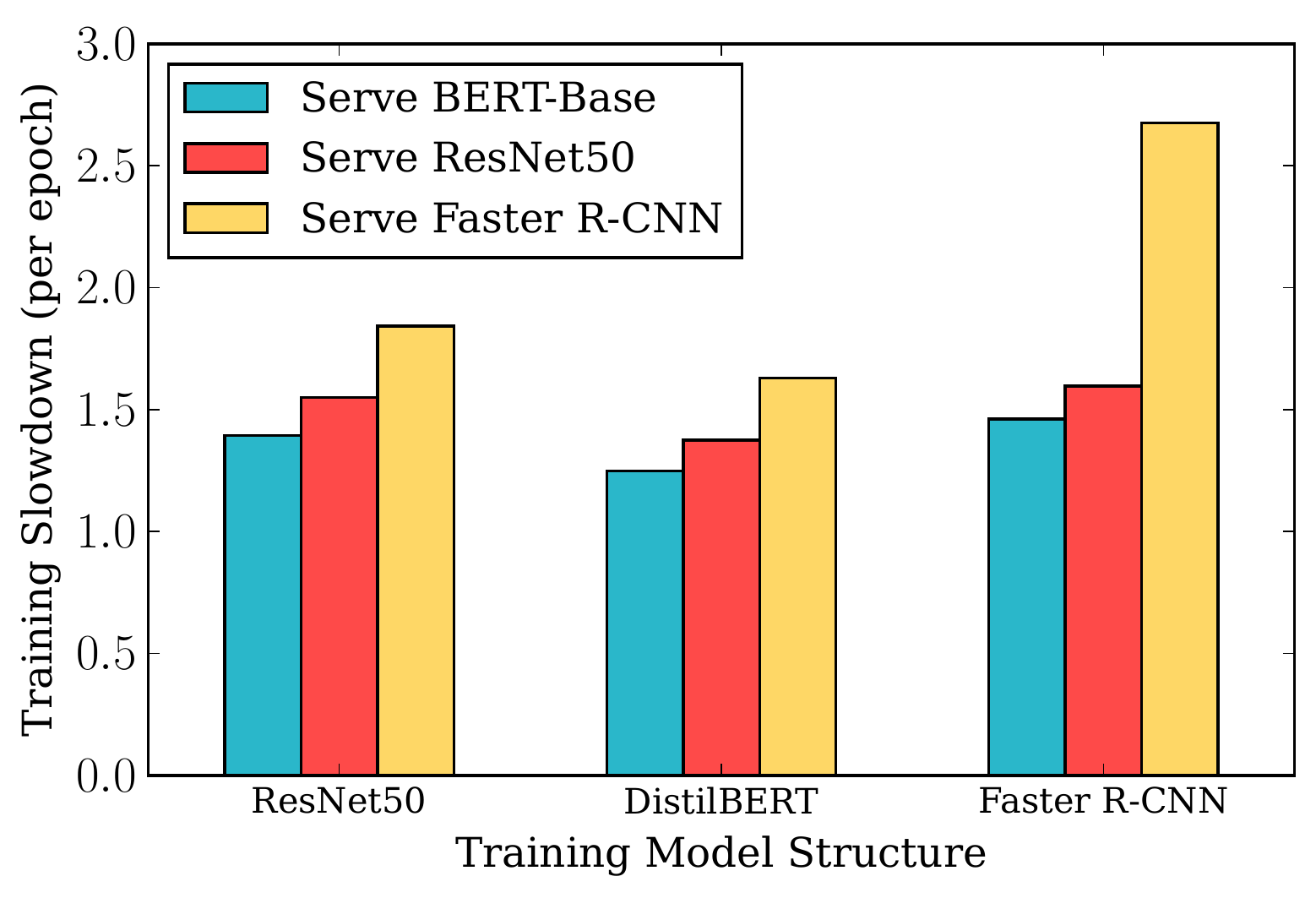}
    \subcaption{Training slowdown.}
\end{subfigure}
\caption{The CDF of ResNet50 inference latency (a) and the slowdown ratios of CL training jobs (b).}
\label{fig:slowdown}
\end{figure}

To housekeep these CL services, our system realizes a dedicated management mechanism with a web interface for multiple roles working together. As illustrated in Figure \ref{fig:panel}, after deployment, a CL service' status is clearly shown in the region \Circled{1}. The information includes the total number of learned classes, current accuracy, the data drift magnitude of service, etc., which provide team members insights about designing better services. Meanwhile, all members can check the model updating history as shown in region \Circled{2}. The history contains every update, no matter it is successfully deployed or rejected to be online since the service is created. Furthermore, they can review detailed information by clicking the task name and then go to the mode card as shown in region \Circled{3}. The model card introduces four kinds of information, including a CL task's detail (e.g., loss function and CL scenario), a benchmarking report, which records the model performance (e.g., accuracy) and CL-related metrics (e.g., model size efficiency), a statistic of the training data and SQL queries generated by the system (shown in \Circled{4}) for users to crawl data and reproduce results, and a validation section designed for quality assurance shown in \Circled{5}. In general, the system considers crucial needs from both researchers and engineers, facilitates collaborative work on improving many CL services, and accelerates the model updating process.

\begin{figure}
  \centering
  \includegraphics[width=1.0\linewidth]{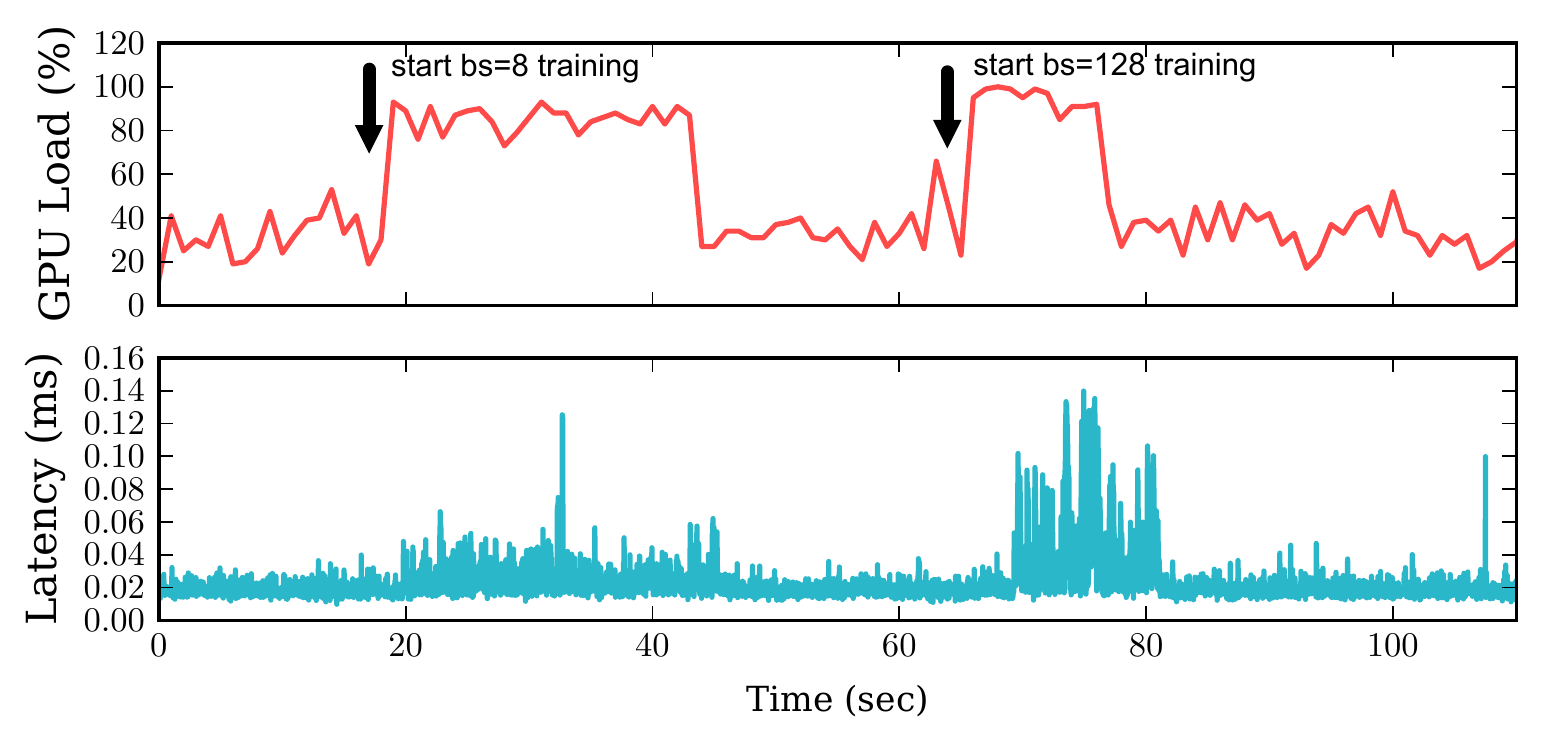}
  \caption{GPU load (top) and inference latency trace (bottom).}
  \label{fig:trace}
\end{figure}

\textbf{Evaluation 2: Preliminary studies about the interference of running training and inference jobs simultaneously in the same cluster.} We first employ ResNet50 IR as our CL service, and simulate users sending requests at a certain rate (100 images/sec). Meanwhile, the system starts updating jobs with batch size 8 for different models (e.g., Faster R-CNN and DistilBERT). We then examine their impact on inference tail latency and training speed. As Figure \ref{fig:slowdown} (a) shows, updating models will increase the 95th percentile tail latency significantly (e.g., over 3x while training a Faster R-CNN model). Also, as shown in Figure \ref{fig:slowdown} (b), due to the inference workload, the training time per epoch has increased by a factor of 1.31 to 2.74, indicating a severe interference. We further investigate their impact on GPU utilization. As Figure \ref{fig:trace} shows, the inference job does not fully utilize the dedicated GPU. On the other hand, adding updating jobs can increase the utilization significantly but it is clear that we need a good scheduling algorithm so that inference latency will not be affected. \textit{We are working on designing a deep reinforcement learning-based method to address the issue.}

\section{Conclusions}
Nowadays, more applications are relying on deep learning (DL) serving systems to provide efficient inference services. Current DL serving systems lack convenient mechanisms to continuously update ML models to deal with the concept drift issue. We introduce ModelCI-e to enable easy continual learning (CL) for model updating in production ML serving systems. It provides a lightweight backend for researchers and engineers to develop CL services, as well as a unified web interface for them to work closely on service management. We believe ModelCI-e can narrow the gap between CL research and model serving in ML production systems.

\bibliographystyle{ACM-Reference-Format}
\bibliography{modelci_e}

\end{document}